\documentclass[letterpaper,showpacs,preprintnumbers,superscriptaddress,aps,nofootinbib,preprint,10pt]{revtex4}
\usepackage{amssymb}
\usepackage[centertags]{amsmath}
\usepackage{txfonts}\usepackage{epsfig}
\usepackage{bm}
\usepackage{color}
\usepackage{graphicx,graphics}
\usepackage{multirow}
\usepackage{float}
\usepackage{ulem}
\usepackage{setspace}
\usepackage{slashed}
\usepackage{booktabs}
\allowdisplaybreaks[4]

\begin{document}

\title{A non-perturbative approach to the study of confinement-deconfinement phase transition in $SU(3)$ dual QCD formulation}

\author{Garima Punetha and H.C. Chandola } \footnote{garimapunetha@gmail.com(corresponding author)}

\affiliation{Centre of Advanced Study, Department of Physics, Kumaun University, Nainital-263001, India}

\begin{abstract}
   We provide an analytical derivation of the confinement deconfinement phase transition  to QCD in the pure gluon sector at finite temperature within the framework of $SU(3)$ dual QCD formulation. The mechanism of color confinement has been explained by adopting magnetic symmetry to extract magnetic monopoles in a gauge invariant way and produce magnetic condensation of $SU(3)$ dual QCD vacuum which guarantees the dual Meissner effect. The deconfinement phase transition has been described by constructing the effective potential at finite temperature in the imaginary time formalism. The magnetic monopole condensate acts as an order parameter for both the confinement deconfinement phase and has been observed to be associated with the appearance/disappearance of magnetic monopole condensate. The evidence of the confinement deconfinement phase transition is further confirmed by the appearance/disappearance of the dual Meissner effect and the existence/breaking of flux tube with temperature. The gauge-invariant  vector and scalar glueball masses at finite temperature have been obtained from the minimization of the effective potential at finite temperature which seems to demonstrate a first order deconfinement phase transition associated with the restoration of magnetic symmetry. 

\end{abstract}


\maketitle

\section{Introduction}\label{S:introduction}
Quantum Chromodynamics (QCD) has been well illustrated as a fundamental theory of strong interaction among quarks and gluons and in consideration distinct notion about the investigation of QCD phase diagram has been performed. The conception of the QCD phase diagram at non-zero temperature ($T$) and chemical potential ($\mu_{B}$) has been one of the eminent and puzzling issue in particle physics \cite{Fukushima, Philipsen}. In general, the spectrum of QCD bound states in vacuum allows a substantial and intriguing phase structure of QCD. It has been well established that the confinement phase in QCD is determined at low $T$ and small $\mu_{B}$ region, while at a high $T$ and $\mu_{B}$ region it approaches towards the deconfinement phase \cite{Fister, Herbst, Hass}. Since the standard perturbative method does not work at the strong coupling regime, as a result, the perturbative technique seems to be inconsistent with the investigation into the QCD phase diagram in the $T-\mu_{B}$ plane. The examination of the QCD running coupling constant at low temperature signify that quarks are firmly bound into hadrons demonstrating a linearly rising quark potential for macroscopic distance. However, at high temperature the thermal fluctuations among hadrons increases which in turn screens the quark confining potential and may be conjectured to undergo as a phase transition from the confined phase to the deconfined phase of hadronic matter. In order to analyze the complex phase structure of QCD at non-zero $T$ and $\mu_{B}$, several effective  models have been designed based on the QCD Lagrangian and symmetries \cite{Hell, Radzhabov, Sakai}.  Unfortunately, detailed study of the phases confronted at non-zero $T$ and $\mu_{B}$ have been eminently model dependent and thus so far have not been examined with first principle QCD calculations. The most familiar approach to deal with the non-perturbative concern of QCD phase diagram is the lattice QCD technique  and that has been turning out to be appropriately favorable for the case of vanishing $\mu_{B}$ where one determine a critical temperature which designate the confinement deconfinement phase transition \cite{Cheng, Aoki(2009), Datta, Borsanyi}. However, the lattice QCD technique has certain complexities associated with the study of non-perturbative phenomena at non-zero $T$ and $\mu_{B}$ as a consequence undergo the infamous sign problem. Moreover, the understanding of  the confinement deconfinement phase transition have been intensely investigated by a variety of non-perturbative field theoretical tools such as functional methods \cite{Fisher, Braun}. In particular, the analysis of the QCD phase boundary at finite $T$ and $\mu_{B}$ serve as one of the leading inquest for a preferred understanding of heavy-ion collision experiments. The relativistic heavy-ion facilities, i.e., the Relativistic Heavy-Ion Collider (RHIC) at BNL, the Large Hadron Collider (LHC) at CERN, and the future Nuclotron-based Ion Collider fAcility (NICA) at JINR, Dubna-Russia contributes to the distinct resolution of the confinement deconfinement phase transition of QCD \cite{Ratti}. Within such reach theoretical studies has been comprehended to investigate the QCD phase diagram and to formulate some specified criterion based on the first principle accompanied by very sharp changes in order parameters for the study of confinement deconfinement phase transition \cite{Andrei, McLerran, Ariel, Fischer(2009)}.

 In the present paper we generally focus on the confinement deconfinement phase transition in the pure gluon sector in the theoretical framework of $SU (3 )$ dual QCD formalism. Both confinement and deconfinement phase transition are treated on equal footing and studied through the mechanism of dual Meissner effect which produces the condensation of dual QCD vacuum. The disappearance  of the dual Meissner effects and the breaking of the flux tube picture has been observed in the deconfinement phase. The deconfinement phase transition seems to be discontinuous demonstrating the order to be first and has been shown to be associated with the restoration of magnetic symmetry.

\section{SU(3) Dual QCD Formulation}
The $SU(3)$ dual QCD formulation is distinct from the typical QCD, i.e. the gauge potential may be expressed in terms of the local as well as topological degrees of freedom \cite{Cho(1980), Cho(1981), pandey, hccgp2016, gphccepl, gphccepj2018, gphccepj2019, gphccappb2019}. The local degrees of freedom remains unrestricted, while the topological degrees of freedom may be characterized by two internal killing vectors, $\hat m$ and $\hat m^{'}$ respectively. With the introduction of $\hat m$ and $\hat m^{'}$, the gauge potential may be given in the following form,
\begin{equation}
{\bf W}_ \mu \, =\,  A_ \mu  \, \hat m + \,  A_ \mu^{'}  \, \hat m^{'}- {g^{-1}}\,
(\, \hat m \times \partial _ \mu \, \hat m)\, - {g^{-1}}\,
(\, \hat m^{'} \times \partial _ \mu \, \hat m^{'}),
\label{2.1}
\end{equation}
where, $ A_\mu$ and $ A_\mu^{'}$ are the $\lambda_3$ and $\lambda_8$- like unrestricted electric  component of the gauge potential. In addition, the topological structure of the gauge symmetry may be described by the homotopy of the mapping, $\Pi_2(G/H) \rightarrow \Pi_2 (SU(3)/U(1)\otimes U^{'}(1)) $, so that the insertion of the magnetic symmetry distributes out the topological degrees of freedom into the dynamics explicitly.  Let us adopt the subsequent parametrization given in the following form,  
\begin{equation}
U= exp\biggl[-\beta^{'} (-\frac{1}{2}t_{3}+\frac{1}{2}\sqrt{3}t_{8})\biggr] \times e^{-\alpha t_{n}} exp\biggl[-(\beta-\frac{1}{2}\beta^{'}) t_{3}e^{-\alpha t_{2}}\biggr], (\beta= n\varphi, \beta^{'}= n^{'}\varphi),
\label{2.2}
\end{equation} 
such that $\hat{m}$ consists of $i$ and $u$- spin subgroup pursued by $n$ and $n^{'}$ windings of $SU(3)$ gauge group given in the following form, 
\begin{equation}
\hat m = \left( \begin{array}{c} sin \alpha cos\frac{1}{2} \alpha cos[(\beta-\beta^{'})]\\sin \alpha cos\frac{1}{2}\alpha sin[(\beta-\beta^{'})]\\\frac{1}{4}cos \alpha (3+cos \alpha)\\sin \alpha sin\frac{1}{2}\alpha cos(\beta)\\\             sin \alpha sin\frac{1}{2}\alpha sin(\beta)\\\frac{-1}{2}sin \alpha cos \alpha cos(\beta^{'})\\\frac{-1}{2}sin \alpha cos \alpha sin(\beta^{'})\\\frac{1}{4}\sqrt{3}cos \alpha (1-cos \alpha)      \end{array} \right).
\label{2.3}
\end{equation}
The obtained gauge potential after the above-mentioned parametrization may thus be expressed in the following form,
\begin{equation}
{\bf W_{\mu}} {\buildrel U \over
\longrightarrow} \frac{1}{g}\biggl[\biggl((\partial_{\mu}\beta-\frac{1}{2}\partial_{\mu}\beta^{'})cos\alpha\biggr)\hat \xi_{3} + \frac{1}{2}\sqrt{3}(\partial_{\mu}\beta^{'}cos\alpha)\hat \xi_{8}\biggr],
\label{2.4}
\end{equation}
where the two internal killing vectors $\hat m$ and $\hat{m}^{'}$ becomes the space-time independent $\hat {\xi_{3}}$ and $\hat {\xi_{8}}$ component respectively.

The equivalent field strength may then takes the following form given below, 
$$
{\bf G_{\mu\nu}}{\buildrel U \over
\longrightarrow} -\frac{1}{g}\biggl[sin\alpha\biggl((\partial_{\mu}\alpha \partial_{\nu}\beta -\partial_{\nu}\alpha\partial_{\mu}\beta)-\frac{1}{2}(\partial_{\mu}\alpha \partial_{\nu}\beta^{'} -\partial_{\nu}\alpha\partial_{\mu}\beta^{'})\biggr)\hat m 
$$
\begin{equation}
+ \frac{\sqrt{3}}{2}sin\alpha (\partial_{\mu}\alpha \partial_{\nu}\beta^{'} -\partial_{\nu}\alpha\partial_{\mu}\beta^{'})\hat m^{'}\biggr].
\label{2.5}
\end{equation}
The Lagrangian for the $SU(3)$ dual QCD formulation in addition may be formulated by using the regular dual magnetic potential $B_{\mu}^{(d)}$, $B_{\mu}^{'(d)}$ and introducing the  complex scalar fields $\phi(x)$, $\phi^{'}(x)$ that originate the topological singularities of $\hat{m}$ and $\hat{m}^{'}$ and thus may be expressed in the following form,
\[
\mathcal{L} = -\frac{1}{4}F_{\mu\nu}^{2}-\frac{1}{4}F_{\mu\nu}^{' 2}-\frac{1}{4}B_{\mu\nu}^{2}-\frac{1}{4}B_{\mu\nu}^{'2}+ \bar{\psi_{r}}\gamma^{\mu}[i\partial_{\mu}+\frac{1}{2}g(A_{\mu}+B_{\mu})
\]
\[
+\frac{1}{2\sqrt{3}}g(A^{'}_{\mu}+B^{'}_{\mu})]\psi_{r} +\bar{\psi_{b}}\gamma^{\mu}[i\partial_{\mu}+\frac{1}{2}g(A_{\mu}+B_{\mu})+\frac{1}{2\sqrt{3}}g(A^{'}_{\mu}+B^{'}_{\mu})]\psi_{b}+ 
\]
\[
\bar{\psi_{y}}\gamma^{\mu}[i\partial_{\mu}-\frac{1}{\sqrt{3}}g(A^{'}_{\mu}+B^{'}_{\mu})]\psi_{y} +|(\partial_{\mu}+i\frac{4\pi}{g}(A_{\mu}^{(d)}+B_{\mu}^{(d)}))\phi|^{2}+
\]
\begin{equation}
 |(\partial_{\mu}+i\frac{4\pi\sqrt{(3)}}{g}(A_{\mu}^{'(d)}+B_{\mu}^{'(d)}))\phi^{'}|^{2}- m_{0}(\bar{\psi_{r}}\psi_{r}+\bar{\psi_{b}}\psi_{b}+\bar{\psi_{y}}\psi_{y}),
\label{2.6}
\end{equation}
where, $F_{\mu\nu}$, $F_{\mu\nu}^{'}$, $B_{\mu\nu}$, $B_{\mu\nu}^{'}$ are the abelian field strengths corresponding to the potentials $A_{\mu}$, $A_{\mu}^{'}$, $B_{\mu}^{(d)}$, $B_{\mu}^{'(d)}$ respectively and $A_{\mu}^{(d)}$, $A_{\mu}^{'(d)}$, $B_{\mu}, B_{\mu}^{'}$ are the singular dual potential of the fields  $F_{\mu\nu}^{(d)}$, $F_{\mu\nu}^{'(d)}$, $B_{\mu\nu}^{(d)}$, $B_{\mu\nu}^{'(d)}$ respectively.

The $SU(3)$ dual QCD Lagrangian in absence of color electric sources may be reduced in the following form, 
\begin{equation}
\pounds= -\frac{1}{4}B_{\mu\nu}^{2}-\frac{1}{4}B_{\mu\nu}^{' 2}
+ |(\partial_{\mu}+i\frac{4\pi}{g}B_{\mu}^{(d)})\phi|^{2}+|(\partial_{\mu}+i\frac{4\pi\sqrt{(3)}}{g}B_{\mu}^{'(d)})\phi^{'}|^{2} - V,
\label{2.7}
\end{equation}  
where $V$ generates the dynamical magnetic condensation of the monopoles and establish the confinement of colored flux tube in the dual QCD vacuum. In the strong coupling limit, at the one loop level, the Coleman-Weinberg \cite{coleman73} type effective potential is preferred, however, from the phenomenological point of view, the familiar Quadratic potential is naturally desired and is given in the following form, 
\begin{equation}
V=\frac{48\pi^{2}}{g^{4}}\lambda(\phi^{\ast}\phi-\phi_{0}^{2})^{2}+ \frac{432\pi^{2}}{g^{4}}\lambda^{'}(\phi^{\ast}\phi^{'}-\phi_{0}^{'2})^{2},
\label{2.8}
 \end{equation}
where $\phi_{0}$ and $\phi^{'}_{0}$ are the non-zero vacuum expectation values of the fields $\phi$ and $\phi^{'}$ respectively. The above-mentioned potential recovers the same mass ratios for $\lambda=\lambda^{'}=1$, therefore, is more favorable then the Coleman-Weinberg potential. A comparison of the Coleman-Weinberg potential and Quadratic potential have been presented in figure 1. 
\begin{figure} 
\includegraphics[width=.4\textwidth,origin=c,angle=360]{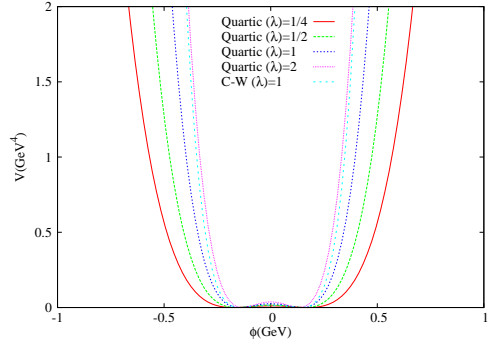}
\caption{(Color online.) A comparison of the Coleman-Weinberg potential and Quadratic potential.}
\label{Figure1}
\end{figure}
The physical spectrum of the magnetically condensed $SU(3)$ dual QCD vacuum contains the magnetic glueballs, two scalar glueballs $m_{\phi}$, $m_{\phi}^{'}$ and two vector glueballs $m_{B}$, $m_{B}^{'}$ respectively. The quadratic potential (\ref{2.9}) fixes the mass ratios as $\frac{m_{\phi}^2}{m_{B}^2}= \frac{3}{2\pi\alpha_s}\lambda$ and $\frac{m_{\phi}^{'2}}{m_{B}^{'2}}= \frac{9}{2\pi\alpha_s}\lambda^{'}$, where $\alpha_s$ is the fine structure constant. The magnetic glueball masses may further be estimated by evaluating the string tension of the flux tube generated by the dynamical breaking of magnetic symmetry. 
 
Further analyzing  the flux tube structure of the $SU(3)$ dual QCD vacuum and using the effective potential (\ref{2.8}), the equation of motion associated with the Lagrangian (\ref{2.7}) are derived in the following form,
\begin{equation}
B_{\mu\nu ,}^{\,\,\,\,\,\,\, \nu} + i
\frac{4\pi}{g}(\phi^*\, {\buildrel \leftrightarrow \over
\partial_\mu}\, \phi)- \frac{32\pi^{2}}{g^{2}} B_\mu^{(d)}\, \phi\,
\phi^* =0\,,
\label{2.9}
\end{equation}
\begin{equation}
(\partial^\mu - i\frac{4\pi}{g} B^{(d)\mu}) (\partial_\mu
  +i\frac{4\pi}{g} B_\mu^{(d)}) \phi + \frac{96\lambda\pi^{2}}{g^{4}}(\phi^*
\phi-\phi_{0}^{2})\phi =0,
\label{2.10} 
\end{equation}
\begin{equation}
(\partial^\mu - i\frac{4\pi\sqrt{3}}{g} B^{' (d)\mu}) (\partial_\mu
  +i\frac{4\pi\sqrt{3}}{g} B_{\mu}^{' (d)}) \phi^{'} + \frac{864\pi^{2}}{g^{4}}\lambda^{'}(\phi^{'*}\phi^{'} -\phi^{' 2}_0)\phi^{'} =0,
\label{2.11}
\end{equation}
\begin{equation}
B_{\mu\nu ,}^{'\,\,\, \nu} + i\frac{4\pi\sqrt{3}}{g}(\phi^{'*}\, {\buildrel \leftrightarrow \over
\partial_\mu}\, \phi^{'})- \frac{96\pi^{2}}{g^2} B_\mu^{' (d)}\, \phi^{'}\,
\phi^{'*} =0\,.
\label{2.12}
\end{equation}
 
where equation (\ref{2.9}), (\ref{2.10}) corresponds to the $\lambda_3$-component and   (\ref{2.11}), (\ref{2.12}) corresponds to the $\lambda_8$-component respectively.

The magnetic condensation of QCD vacuum leads to a definite flux tube structure to the dual QCD vacuum, therefore it is naturally desired to analyze the flux tube structure and the nature of the magnetically condensed vacuum. Using the cylindrical symmetry with the co-ordinates $( \rho,\, \varphi, z ) $ and the flux tube orientation along the direction of $ z $-axis, the dual gauge field and the monopole field for the $\lambda_{3}$ and $\lambda_{8}$ components may be expressed in the following form,
\begin{equation}
B_{\mu}^{(d)}\,\,=\,\, \frac{1}{g}cos\alpha(\partial_{\mu}\beta)\hat{m},\,\, \phi (x)\,=\, exp\,(i\,n\,\varphi)\,\chi(\rho),\,\,\,\,(n\,=\,0,\,\pm\, 1,\, \cdots),
\label{2.13}
\end{equation}
\begin{equation}
B_\mu^{' (d)}\,\,=\,\,\frac{\sqrt{3}}{2g}\,cos \,\alpha\, (\partial_\mu\,\beta^{'})\,\hat m^{'},\,\, \phi^{'} (x)\,=\, exp\,(i\,n^{'}\,\varphi)\,\chi^{'}(\rho),\,\,\,\,(n^{'}\,=\,0,\,\pm\, 1,\, \cdots),
\label{2.14}
\end{equation}
and the equation of motion reduces to the following form,
\begin{equation}
\frac {d}{d \rho}\biggl [\frac{1}{\rho}\, \frac {d}{d \rho}\biggl (\rho
B(\rho) \biggr ) \biggr ]\,-\,\biggl( \frac{16\pi}{\alpha_s}\biggr)^{1/2} \biggl
(\frac {n}{\rho}\,+
\, \biggl(\frac{4\pi}{\alpha_s}\biggr)^{1/2} B(\rho) \biggr)\, \chi^2
\,(\rho)\,=\,0,
\label{2.15}
 \end{equation}
\begin{equation}
\frac {1}{\rho}\, \frac {d}{d \rho} \biggl (\rho\, \frac {d
\chi(\rho)}{d
   \rho}\biggr )\,-\,\biggl [ \biggl (\frac {n}{\rho}\,+\, \biggl(\frac{4
\pi}{\alpha_s}\biggr)^{1/2} B(\rho) \biggr )^2 \,+
\,\frac{6\lambda}{\alpha_s^{2}} \biggl (
 \chi^2 -\phi_0^2 \biggr )\biggr ]\, \chi (\rho) =0,
 \label{2.16}
\end{equation}
 \begin{equation}
\frac {d}{d \rho}\biggl [\frac{1}{\rho}\, \frac {d}{d \rho}\biggl (\rho
B^{'}(\rho) \biggr ) \biggr ]\,-\,\biggl(\frac{48\pi}{\alpha_s}\biggr)^{1/2} \biggl
(\frac {n^{'}}{\rho}\,+
\, \biggl(\frac{12\pi}{\alpha_s}\biggr)^{1/2} B^{'}(\rho) \biggr)\, \chi^{'2}
\,(\rho)\,=\,0,
\label{2.17}
 \end{equation}
\begin{equation}
\frac {1}{\rho}\, \frac {d}{d \rho} \biggl (\rho\, \frac {d
\chi^{'}(\rho)}{d \rho}\biggr )\,-\,\biggl [ \biggl (\frac {n^{'}}{\rho}\,+\, \biggl(\frac{12
\pi}{\alpha_s}\biggr)^{1/2} B(\rho^{'}) \biggr )^2 \,+
\,\frac{54\lambda^{'}}{\alpha_s^{2}} \biggl (
 \chi^{'2} -\phi^{' 2}_0 \biggr )\biggr ]\, \chi^{'} (\rho) =0.
\label{2.18}
 \end{equation}
Using the Lagrangian equation (\ref{2.7}), the energy per unit length carried by the associated flux tube configuration governed by the field  equations (\ref{2.15}), (\ref{2.16}), (\ref{2.17}) and (\ref{2.18}) may be obtained in the following form,  
\[
k\,(B,\,\chi, B^{'}, \chi^{'})\,=\, 2\pi  \int_0^\infty \rho \, d \rho \Biggl [ \frac {1}{2\rho^2}  \Biggl (\frac {d}{d \rho} (\rho \, B(\rho)  ) \Biggr )^2  +\Biggl (\frac {d}{d \rho}\, \chi (\rho) \Biggr )^2 +\Biggl (\frac {4\pi}{g} B(\rho) + \frac {n}{\rho} \Biggr )^2 \chi^2 (\rho)
\]
\[
 + \frac{3\lambda}{\alpha_{s}^{2}}(\chi^{2}-\phi_{0}^{2})^{2} \Biggr ]+  2\pi  \int_0^\infty \rho \, d \rho \Biggl [ \frac {1}{2\rho^2}  \Biggl (\frac {d}{d \rho} (\rho \, B^{'}(\rho)  ) \Biggr )^2 +\Biggl (\frac {d}{d \rho}\, \chi^{'} (\rho) \Biggr )^2
\]
\begin{equation}
  +\Biggl (\frac {4\pi\sqrt{3}}{g} B^{'}(\rho) + \frac {n^{'}}{\rho} \Biggr )^2 \chi^{' 2} (\rho) + \frac{27\lambda^{'}}{\alpha_{s}^{2}}(\chi^{' 2}-\phi^{' 2}_{0})^{2} \Biggr ].
\label{2.20}
\end{equation}
The acceptable boundary condition for the large scale behavior of QCD is obtained in the asymptotic limit, $B(\rho)\stackrel {\rho \to \infty} {\longrightarrow} -\frac{ng}{4\pi\rho}$, $\phi \stackrel {\rho \to \infty} {\longrightarrow} \phi_{0}$, $B^{'}(\rho)\stackrel {\rho \to \infty} {\longrightarrow} -\frac{n^{'}g}{4\sqrt{3}\pi\rho}$ and $\phi^{'} \stackrel {\rho \to \infty} {\longrightarrow} \phi^{'}_{0}$, which leads to the asymptotic solution for $B(\rho)= -\frac{ng}{4\pi\rho}[1+F(\rho)]$ and $B^{'}(\rho)= -\frac{n^{'}g}{4\sqrt{3}\pi\rho}[1+G(\rho)]$, where the function $F(\rho)$ and $G(\rho)$ are obtained in the following form 
\begin{equation}
F(\rho)\,\,\,\stackrel {\rho \to \infty} {\longrightarrow}\,\,\,-\,n\,\,+\,\,C\sqrt \rho/exp\, (m_B \,
\rho), \,\,\, G(\rho)\,\,\,\stackrel {\rho \to \infty} {\longrightarrow}\,\,\,-\,n^{'}\,\,+\,\,C\sqrt \rho/exp\, (m_B^{'} \,
\rho).
\label{2.21}
\end{equation}
The resulting energy per unit length of the corresponding flux tube takes the following form,
$$
k_q\, = \, 2\pi\, \int_0^\infty \, \rho\, d \rho \Biggl [\frac {n^2 g^2}{32\pi^2\rho^2}\Biggl( \frac{d F}{d \rho} \Biggr )^2 \,+\, \frac {n^2}{\rho^2}\, F^2 (\rho)\chi^2 (\rho)+ \Biggl (\frac{d \chi}{d \rho} \Biggr )^2+ \frac{3\lambda}{\alpha_{s}^{2}}(\chi^{2}-\phi_{0}^{2})^{2}  \Biggr ]
$$
\begin{equation}
+ 2\pi\, \int_0^\infty \, \rho\, d \rho \Biggl [\frac {n^{'^2} g^2}{96\pi^2\rho^2}\Biggl( \frac{d G}{d \rho} \Biggr )^2 \,+\, \frac {n^{'2}}{\rho^2}\, G^2 (\rho)\chi^{'2} (\rho)+ \Biggl (\frac{d \chi^{'}}{d \rho} \Biggr )^2+ \frac{27\lambda^{'}}{\alpha_{s}^{2}}(\chi^{'2}-\phi^{' 2}_{0})^{2}  \Biggr ].
\label{2.22}
\end{equation}
The dual dynamics of the QCD vacuum certainly reveals that the physical state has to be color neutral, however, originally there were four magnetic glueballs, amongst them only two are color singlets. Undoubtedly, to eliminate the other two magnetic glueball masses from the physical spectrum, the property of color reflection invariance come into existence. The color reflection invariance determines the hadron spectrum simply by asserting the subsequent conditions, $m_{\phi}= m_{\phi}^{'}= \bar{m_{\phi}}$ and $m_{B}= m_{B}^{'}= \bar{m_{B}}$ or $\phi^{'}_{0}= \frac{1}{\sqrt{3}}\phi_{0}, \lambda_{'}= \frac{1}{3}\lambda$.
With such consideration, the energy per unit length in relationship with the regge slope parameter is expressed in the following form$,
k_{q}= \frac{1}{2\pi\alpha^{'}} = \gamma_{q}\phi_{0}^{2}= \frac{\alpha_{s}}{8\pi}\gamma_{q}\bar m_{B}^{2},
$
where $\gamma_{q}$ is a dimensionless parameter. As a result, using the numerical computation of equation (\ref{2.10}), for $\gamma_{q}$, one may obtain the vector ($\bar m_{B}$) and scalar ($\bar m_{\phi}$) magnetic glueball masses as a function of $\alpha_s$ and these results are obtained in table (\ref{Table:1}) \cite{ChoPRL}.
\begin{table}
\centering
\caption{The masses of vector and scalar glueball as functions of $\alpha_{s}$.}
\label{Table:1}
\begin{tabular*}{\columnwidth}{@{\extracolsep{\fill}}lllllll@{}}
\hline
$\lambda$ & $\alpha_s$ & $\bar m_{\phi}(GeV)$ & $\bar m_{B}(GeV)$ & $\lambda_{QCD}^{(d)}(GeV^{-1})$ & $\xi_{QCD}^{(d)}(GeV^{-1})$& $\kappa_{QCD}^{(d)}$\\ 
\hline
$\frac{1}{4}$ & 0.25 & 1.21 & 1.74 &  0.57 & 0.83  & 0.69 \\
\hline
 $\frac{1}{2}$  & 0.24 & 1.68  & 1.63 &  0.61 & 0.59 & 0.99\\
 \hline
1  & 0.23 & 2.16 & 1.53 & 0.65 & 0.46 &  1.42\\
\hline
2   & 0.22 & 2.89 & 1.42 & 0.70  & 0.34 &  2.05\\ 
\hline
\end{tabular*} 
\end{table}
\section{Deconfinement Phase Transition in $SU(3)$ Dual QCD}
In the present section, we elucidate the mechanism for the deconfinement phase transition at finite temperature in the pure gluon sector and start with the construction of partitional functional for the $SU(3)$ dual QCD vacuum in the following form \cite{Ichie},
\begin{equation}
Z[J]= \int D[\phi]D[B_{\mu}^{(d)}]D[\phi^{'}]D[B_{\mu}^{'(d)}]exp(-i\int d^{4}x(\mathcal{L}_{d}^{(m)}- J|\phi|^{2}-J^{'}|\phi^{'}|^{2})),
\label{3.1}
\end{equation}
here, the quadratic source term has been used which invariably bring $\phi$ and $\phi^{'}$ real even in the negative-curvature region and preserve the symmetry of the classical potential. Separating the $SU(3)$ QCD monopole field $\phi$ and $\phi^{'}$ into its mean field and fluctuation part,
$
\phi\rightarrow (\phi + \tilde\phi)exp(i\xi),\,\,\phi^{'} \rightarrow (\phi^{'} + \tilde\phi^{'})exp(i\xi^{'}),
$ 
the $II^{nd}$ integrand term in equation (\ref{3.1}) is written in the following form,
\[
\mathcal{L}_{d}^{(m)}- J|\phi|^{2}- J^{'}|\phi^{'}|^{2}= \mathcal{L}_{cl}(\phi, \phi^{'})- J\phi^{2}-\frac{1}{4}(\partial_{\mu}B_{\nu}^{(d)}- \partial_{\nu}B_{\mu}^{(d)})^{2}+ \frac{1}{2}m_{B}^{2}(B_{\mu}^{(d)})^{2} 
\]
\[
+ [(\partial_{\mu}\tilde{\phi})^{2}-(m_{\phi}\tilde{\phi})^{2}] +\left[ \frac{4\pi}{\alpha_{s}}(B_{\mu}^{(d)})^{2}(2\phi\tilde{\phi} +\tilde{\phi}^{2})-\frac{3\lambda}{\alpha_{s}^{2}}(\tilde{\phi}^{4}+4\phi\tilde{\phi}^{3}) \right]  
\]
\[
-\left[ \frac{12\lambda}{\alpha_{s}^{2}}(\phi^{2}-\phi_{0}^{2})\phi+2J\phi\right] \tilde{\phi} 
-J^{'}\phi^{' 2}-\frac{1}{4}(\partial_{\mu}B_{\nu}^{'(d)}- \partial_{\nu}B_{\mu}^{'(d)})^{2}
\]
\[
+ \frac{1}{2}m_{B}^{' 2}(B_{\mu}^{' (d)})^{2} +[(\partial_{\mu}\tilde{\phi}^{'})^{2}-(m^{'}_{\phi}
\tilde{\phi}^{'})^{2}]+ [\frac{12\pi}{\alpha_{s}}(B_{\mu}^{' (d)})^{2} (2\phi^{'}\tilde{\phi}^{'}+\tilde{\phi}^{' 2})
\]
\begin{equation}
- \frac{27\lambda^{'}}{\alpha_{s}^{2}}(\tilde{\phi}^{' 4}+4\phi^{'}\tilde{\phi}^{' 3})]- [\frac{108\lambda^{'}}{\alpha_{s}^{2}}(\phi^{' 2}-\phi_{0}^{' 2})\phi^{'}+2J^{'}\phi^{'}]\tilde{\phi}^{'},
\label{3.2}
\end{equation}
where $\mathcal{L}_{cl}(\phi, \phi^{'})$ is the classical part given in the following form, $\mathcal{L}_{cl}(\phi, \phi^{'})= -\frac{3\lambda}{\alpha^{2}}(\phi^2 - \phi^{2}_0)- \frac{27\lambda^{'}}{\alpha^{2}}(\phi^{'2} - \phi^{'2}_0) $.

The masses of the vector and scalar magnetic glueballs are given in the following form,
\[
m_{B}^{2}= \frac{8\pi}{\alpha_{s}}\phi^{2}, \,\,\,\,  m_{\phi}^{2}=\frac{6\lambda}{\alpha_{s}^{2}}(3\phi^{2}-\phi^{2}_{0}) +J,
\]
\begin{equation}
m_{B}^{' 2}= \frac{24\pi}{\alpha_{s}}\phi^{' 2}, \,\,\,\,  m_{\phi}^{' 2}=\frac{54\lambda}{\alpha_{s}^{2}}(3\phi^{' 2}- \phi^{'2}_{0})+J^{'},
\label{3.3}
\end{equation}
here the scalar glueball masses are always real owing to the sources $J$ and $J^{'}$ having relationship with scalar fields $\phi$ and $\phi^{'}$ in the following form, $J= -\frac{6\lambda}{\alpha_{s}^{2}}(\phi^{2}-\phi_{0}^{2})$ and $J^{'}= -\frac{54\lambda^{'}}{\alpha^{2}_{s}}(\phi^{' 2}-\phi_{0}^{' 2})$. The vector and scalar field in the above-mentioned partition functional are integrated out and the expression reduces into the following form,
\[
Z[J]= exp[i\int d^{4}x(\mathcal{L}_{cl}(\phi, \phi^{'})-J\phi^{2}- J^{'}\phi^{' 2}]
\]
\[
\times [Det(iD_{\mu\nu}^{-1}({\bf B, k}))]^{-1}[Det(i\bigtriangleup^{-1}(\phi, k))]^{-1/2}\times [Det(iD_{\mu\nu}^{-1}({\bf B^{'}, k}))]^{-1}
\]
\begin{equation}
[Det(i\bigtriangleup^{-1}(\phi^{'}, k))]^{-1/2},
\label{3.4}
\end{equation}
where the respective vector and scalar propagator in the momentum representation of the magnetically condensed $SU(3)$ dual QCD vacuum are given in the following form,
\[
D_{\mu\nu}({\bf B, k})= \frac{i}{k^{2}-m_{B}^{2}+i\epsilon}\biggl( g_{\mu\nu}-\frac{k_{\mu}k_{\nu}}{m_{B}^{2}}\biggr), \,\,\, \Delta(\phi, k)= -\frac{i}{k^{2}-m_{\phi}^{2}+i\epsilon},
\]
\begin{equation}
D_{\mu\nu}({\bf B^{'}, k})= \frac{i}{k^{2}-m_{B}^{' 2}+i\epsilon}\biggl( g_{\mu\nu}-\frac{k_{\mu}k_{\nu}}{m_{B}^{' 2}}\biggr), \,\,\, \Delta(\phi^{'}, k)= -\frac{i}{k^{2}-m_{\phi}^{' 2}+i\epsilon}.
\label{3.5}
\end{equation}
Hence, the effective action is constructed and obtained in the following form, 
\[
S_{eff}(\phi, \phi^{'}) = -ilnZ[J]+ \int d^{4}x J\phi^{2} + \int d^{4}x J^{'}\phi^{'2}
\]
\[
=\int d^{4}x  \mathcal{L}_{cl}(\phi, \phi^{'})+ iln Det(iD_{\mu\nu}^{-1}({\bf B, k}))
\]
\begin{equation}
+ \frac{i}{2}ln Det(i\bigtriangleup^{-1}(\phi,k))+ iln Det(iD_{\mu\nu}^{-1}({\bf B^{'}, k}))+ \frac{i}{2}ln Det(i\bigtriangleup^{-1}(\phi^{'},k)).
\label{3.6}
\end{equation} 
The above-mentioned effective action at finite temperatures substantially reproduces the effective potential at finite temperature which may be expressed in the following form, 
\[
V_{eff}(\phi, \phi^{'})= -\frac{S_{eff}}{\int d^{4}x} = \frac{3\lambda}{\alpha_{s}^{2}}(\phi^{2}-\phi^{2}_{0})^{2} + \frac{27\lambda^{'}}{\alpha_{s}^{2}}(\phi^{' 2}-\phi^{' 2}_{0})^{2}+ 
\]
\[
3\int \frac{d^4 k}{i(2\pi)^{4}}ln(m_{B}^{2}-k^{2}-i\epsilon) + \frac{1}{2}\int \frac{d^4 k}{i(2\pi)^4}ln(m_{\phi}^{2}-k^2- i\epsilon) 
\]
\begin{equation}
+ 3\int \frac{d^4 k}{i(2\pi)^{4}}ln(m_{B}^{' 2}-k^{2}-i\epsilon) + \frac{1}{2}\int \frac{d^4 k}{i(2\pi)^4}ln(m_{\phi}^{' 2}-k^2- i\epsilon).
\label{3.7}
\end{equation}
The corresponding effective potential at finite temperature is extracted by replacing the $k_{0}-$ integration over the infinite sum over the Matsubara frequency which reduces the effective potential in the following form, 
\[
V_{eff}(\phi, \phi^{'},\beta)= \frac{3\lambda}{\alpha_{s}^{2}}(\phi^{2}-\phi_{0}^{2})^{2}+  \frac{27\lambda^{'}}{\alpha_{s}^{2}}(\phi^{' 2}-\phi^{' 2}_{0})^{2}
\]
\[
+ \frac{3}{\beta}\sum_{n=-\infty}^{\infty} \int \frac{d^{3}k}{(2\pi)^{3}}ln[(\frac{2\pi n}{\beta})^{2}+ k^{2}+m_{B}^{2}] + \frac{1}{2\beta}\sum_{n=-\infty}^{\infty} \int \frac{d^{3}k}{(2\pi)^{3}}ln[(\frac{2\pi n}{\beta})^{2}+ k^{2}+m_{\phi}^{2}] 
\]
\begin{equation}
+\frac{3}{\beta}\sum_{n=-\infty}^{\infty} \int \frac{d^{3}k}{(2\pi)^{3}}ln[(\frac{2\pi n}{\beta})^{2}+ k^{2}+m_{B}^{' 2}] + \frac{1}{2\beta}\sum_{n=-\infty}^{\infty} \int \frac{d^{3}k}{(2\pi)^{3}}ln[(\frac{2\pi n}{\beta})^{2}+ k^{2}+m_{\phi}^{' 2}].
\label{3.8}
\end{equation}
The final expression obtained after summation over $n$ and the angular integration leads to the effective potential at finite temperature in the following form,   
\[
V_{eff}(\phi, \phi^{'})=  \frac{3\lambda}{\alpha_{s}^{2}}(\phi^{2}-\phi^{2}_{0})^{2}+ \frac{27\lambda^{'}}{\alpha_{s}^{2}}(\phi^{' 2}-\phi^{' 2}_{0})^{2}
\]
\[
+ 3\frac{T}{\pi^2}\int_0^\infty dk k^2ln(1-e^{-\sqrt{k^2+m_{B}^2}/T}) + \frac{T}{2\pi^2}\int_0^\infty dk k^2ln(1-e^{-\sqrt{k^2+m_{\phi}^2}/T}) 
\]
\begin{equation}
+ 3\frac{T}{\pi^2}\int_0^\infty dk k^2ln(1-e^{-\sqrt{k^2+m^{' 2}_{B}}/T}) + \frac{T}{2\pi^2}\int_0^\infty dk k^2ln(1-e^{-\sqrt{k^2+m^{' 2}_{\phi}}/T}).
\label{3.9}
\end{equation}
In order to formulate the effective potential composed of the absolute physical states, the property of color reflection invariance has been enforced as a residual symmetry of the $SU(3)$ dual QCD vacuum staying intact in the course of the magnetic condensation of the vacuum. The obtained expression for the effective potential at finite temperature is given in the following form,
\begin{equation}
V_{eff}(\phi,T)= \frac{4\lambda}{\alpha^{2}}(\phi^{2}-\phi_{0}^{2})^{2} -\frac{7\pi^2T^4}{45} + \frac{T^2}{12}[\bar{m}_{\phi}^{2}+6\bar{m}_B^{2}].
\label{3.10}
\end{equation}
\section{Numerical Results}
In the present section, the existence of the deconfinement phase transition has been examined by analyzing the above-mentioned effective potential ($V_{eff}(\phi,T)$)(\ref{3.10}), as a function of magnetic monopole condensate ($\phi$), around the critical temperature value for various coupling and have been depicted in figure 2. 
\begin{figure}
\includegraphics[width=.4\textwidth,origin=c,angle=360]{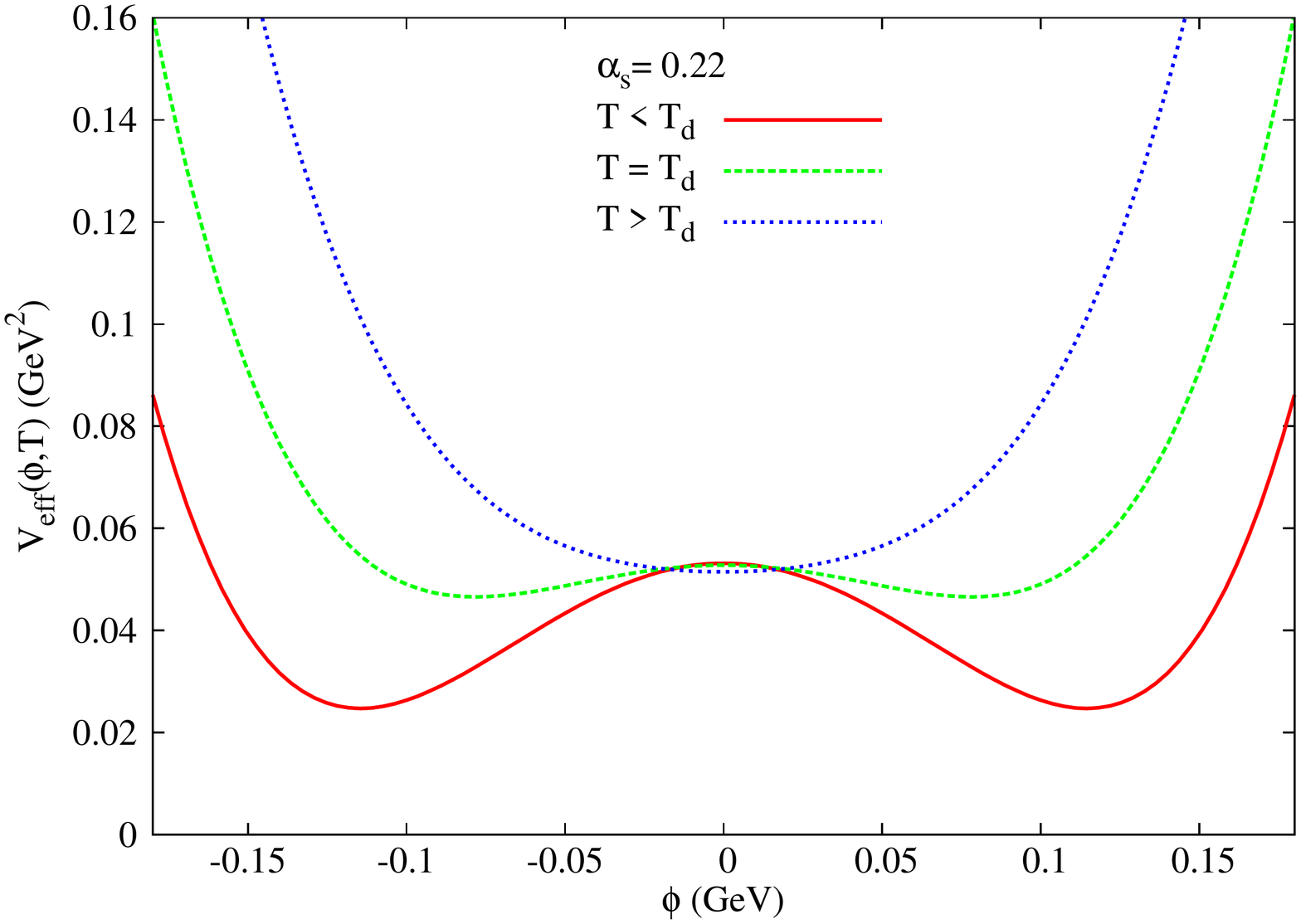}
\includegraphics[width=.4\textwidth,origin=c,angle=360]{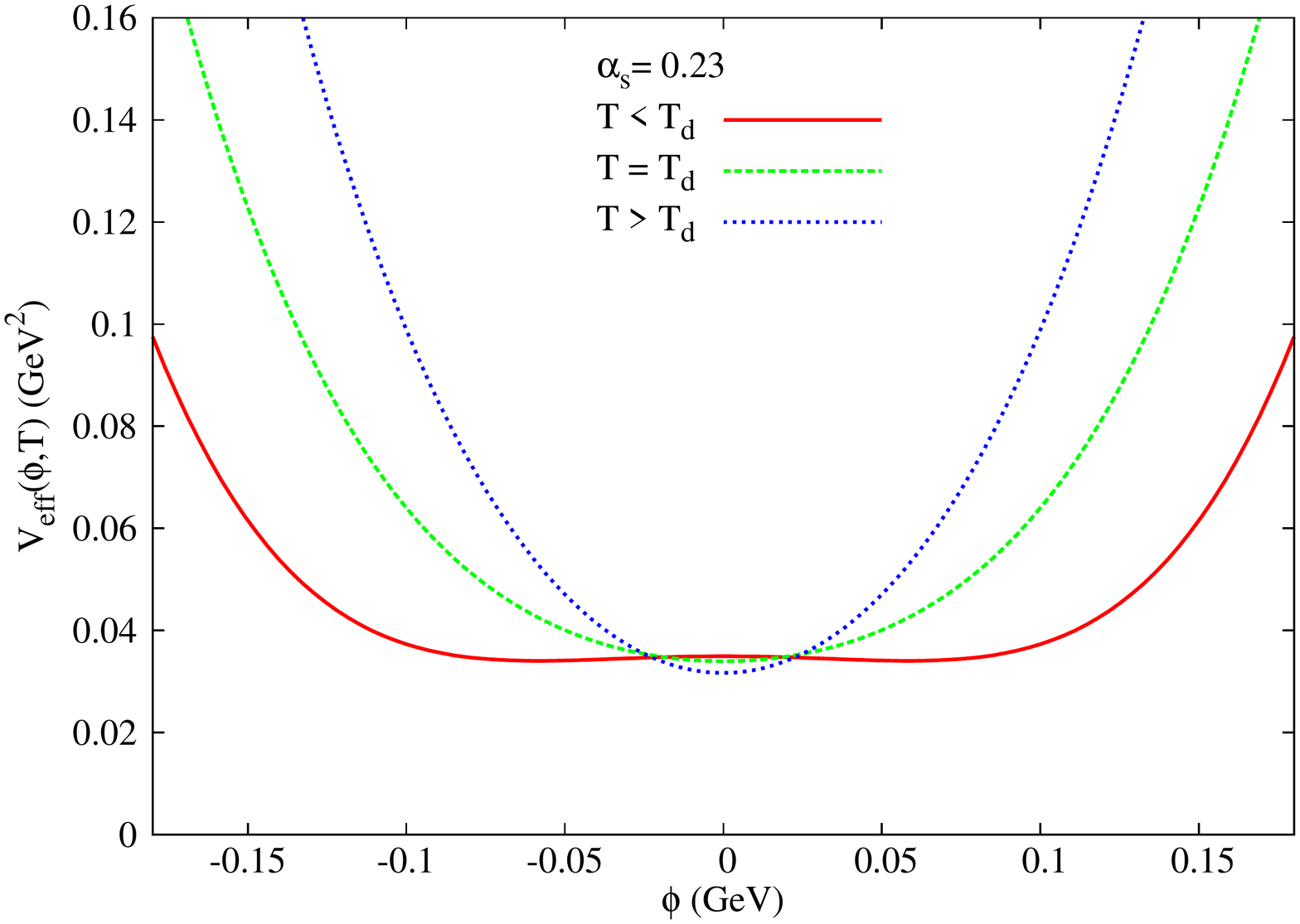}\\
\includegraphics[width=.4\textwidth,origin=c,angle=360]{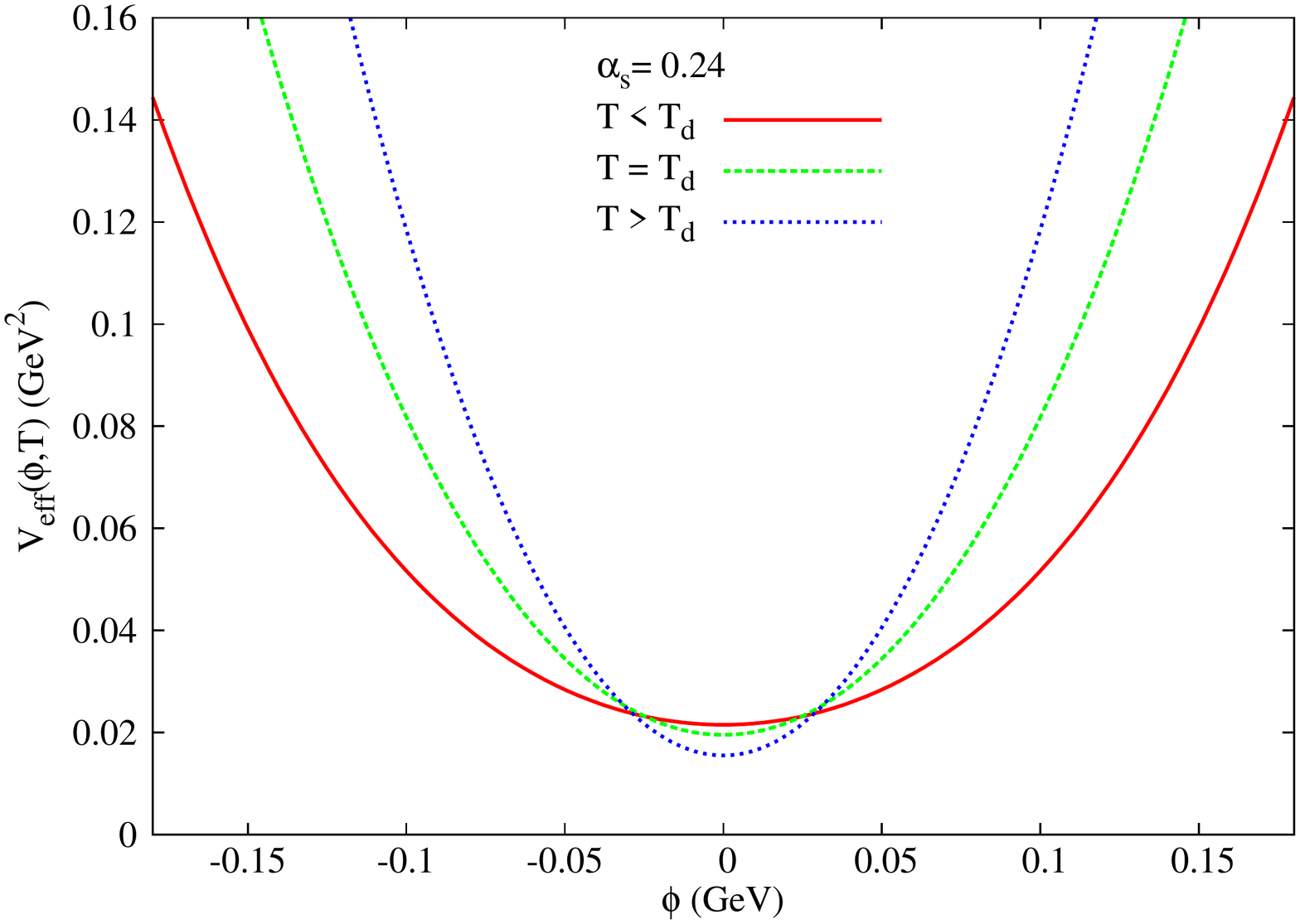}
\includegraphics[width=.4\textwidth,origin=c,angle=360]{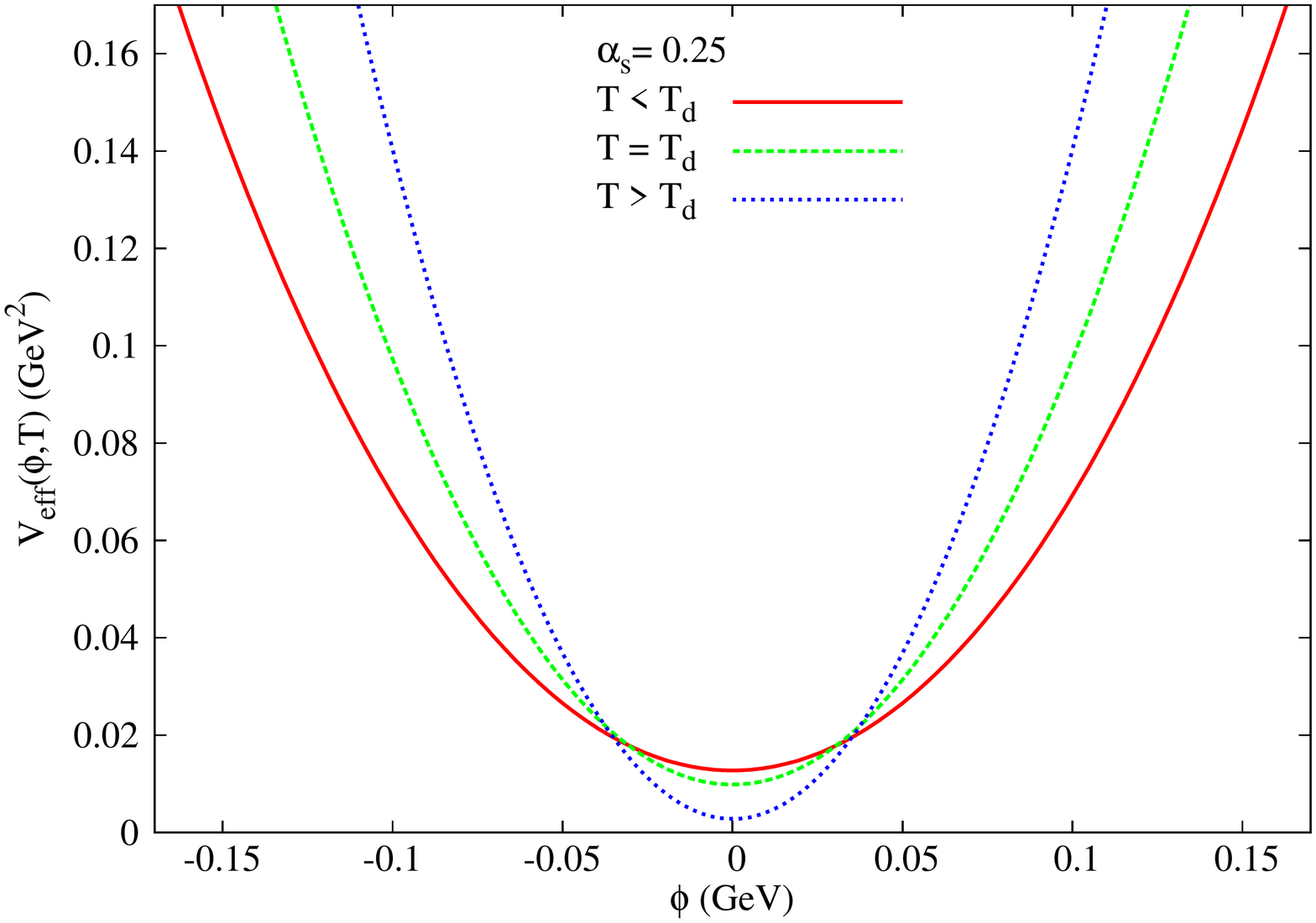}
\caption{(Color online.) The behavior of effective potential at finite temperature with monopole condensate for $\alpha_s$ = $0.22$, $0.23$, $0.24$ and $0.25$ coupling respectively. }
\label{Figure2}
\end{figure}
It demonstrates that for higher temperature ($T>T_d$), the magnetic symmetry tends to be restored and the system moves towards the confinement phase. However below $T_{d}$, the magnetic symmetry is dynamically broken pushing the system towards the confined phase where the local minima of potential corresponds to the physical stable state.
The magnetic monopole condensate acts as an order parameter for explaining the phenomenon of deconfinement phase transition. The same behavior is depicted through the $3D$ behavior of $V_{eff}(\phi, T)$ as a function of the magnetic monopole condensate. The variation indicates the restoration of magnetic symmetry above the critical temperature. The various plots at different coupling have been demonstrated in figure 3.
\begin{figure}
\includegraphics[width=.4\textwidth,origin=c,angle=360]{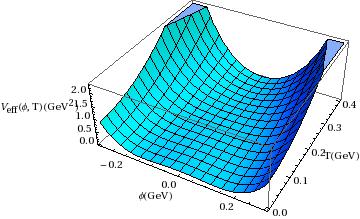}
\includegraphics[width=.4\textwidth,origin=c,angle=360]{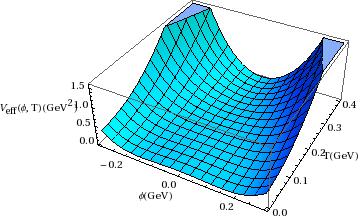}\\
\includegraphics[width=.4\textwidth,origin=c,angle=360]{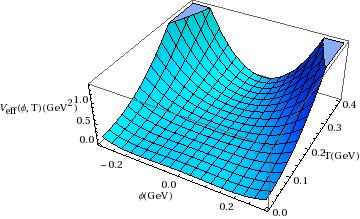}
\includegraphics[width=.4\textwidth,origin=c,angle=360]{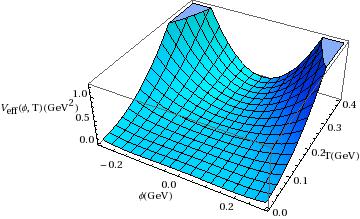}
\caption{(Color online.) The behavior of effective potential at finite temperature  with monopole condensate and temperature for $\alpha_s$ = $0.25$, $0.24$, $0.23$ and $0.22$ coupling respectively. }
\label{Figure3}
\end{figure}
The behavior of the vacuum expectation value of the QCD monopole condensate as a function of the temperature in the $SU(3)$ dual QCD vacuum, obtained after the minimization of the effective potential (\ref{3.10}) at finite temperature has been plotted in figure 3. The contour plots for the thermal response of the $<\phi>_{0}^{(T)}$  with temperature for various coupling have also been presented in figure 4. One finds $<\phi>_{0}^{(T)}$ = $0.174 GeV$, $0.163 GeV$, $0.153 GeV$, $0.142 GeV$ at $T=0$, and further decreases monotonously vanishing near the deconfinement temperature ($T_{d}$) of $0.135 GeV$, $0.170 GeV$, $0.210 GeV$ and $0.243 GeV$ for $\alpha_s$ = $0.25$, $0.24$, $0.23$ and $0.22$ coupling respectively. 
\begin{figure}
\centering
\includegraphics[width=.4\textwidth,origin=c,angle=360]{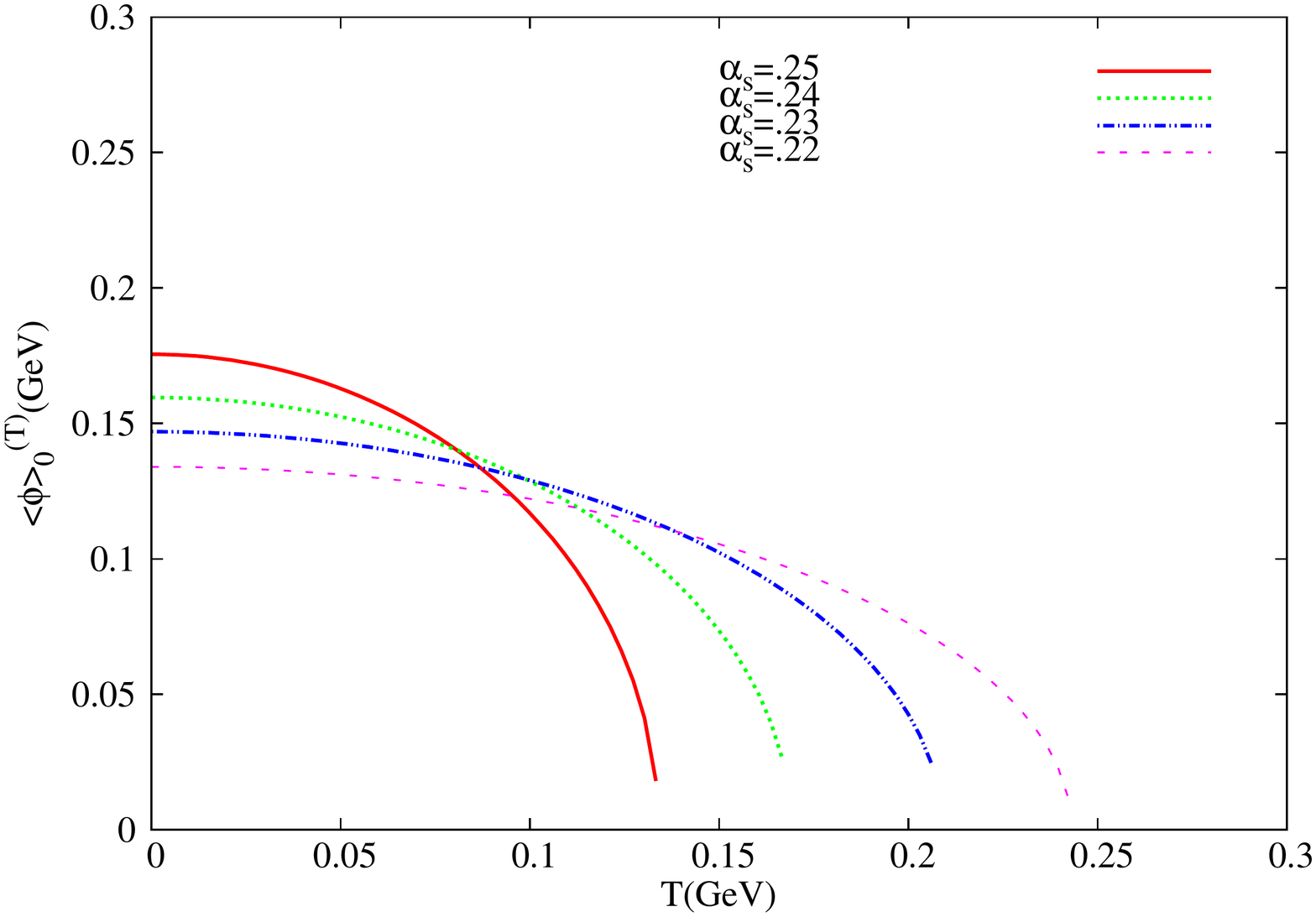}
\caption{(Color online.) Thermal response of the vacuum expectation value of the monopole field with temperature in $SU(3)$ dual QCD vacuum for $\alpha_s$ = $0.25$, $0.24$, $0.23$ and $0.22$ coupling respectively.}
\label{Figure4}
\end{figure}
\begin{figure}
\centering
\includegraphics[width=.4\textwidth,origin=c,angle=360]{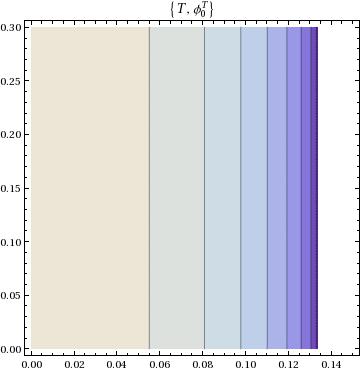}
\includegraphics[width=.4\textwidth,origin=c,angle=360]{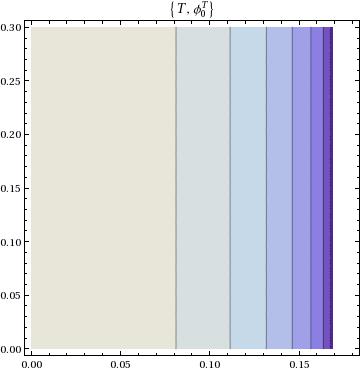}\\
\includegraphics[width=.4\textwidth,origin=c,angle=360]{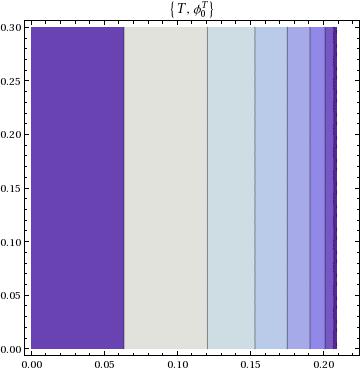}
\includegraphics[width=.4\textwidth,origin=c,angle=360]{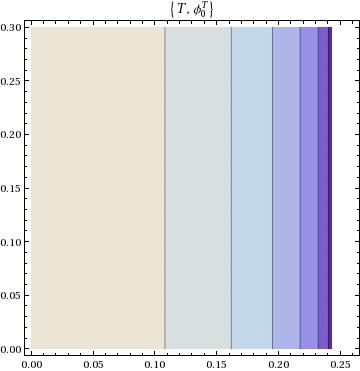}
\caption{(Color online.) Contour plot for the thermal response of the vacuum expectation value of the magnetic monopole field with temperature in $SU(3)$ dual QCD vacuum for $\alpha_s$ = $0.25$, $0.24$, $0.23$ and $0.22$ coupling respectively. }
\label{Figure5}
\end{figure}
Here, we find a large reduction of the $<\phi>_{0}^{T}$ near the deconfinement temperature and gets a first-order deconfinement phase transition. These estimates for the deconfinement temperature seem to be in agreement with the lattice QCD predictions. Moreover, the above idea enables us to investigate the gauge-invariant vector ($\bar{m}_{B}^{(T)}$) and scalar glueball masses ($\bar{m}_{\phi}^{(T)}$) at the finite temperature. Infact, the temperature dependence of the dynamical magnetic glueball masses are obtained from the minimum of effective potential at finite temperature and are given in the following form,
\begin{equation}
\bar{m}_{\phi}^{(T)}= \frac{\sqrt{12\pi}}{\alpha_{s}} \sqrt{\phi^{2}_{0}-\biggl(\frac{4\pi\alpha_{s}+\lambda}{\lambda}\biggr)\frac{T^{2}}{8}}\,, \bar{m}_{B}^{(T)}= \sqrt{\frac{8\pi}{\alpha_{s}}}\sqrt{\phi^{2}_{0}-\biggl(\frac{4\pi\alpha_{s}+\lambda}{\lambda}\biggr)\frac{T^{2}}{8}}.
\label{3.11}
\end{equation}
We next demonstrate the variation of $\bar{m}_{B}^{(T)}$, $\bar{m}_{\phi}^{(T)}$ as function of the temperature $T$  and the contour plot for their thermal response have been presented in figure 5, 6 and 7 respectively. The dynamical magnetic glueball masses at finite temperature drop down to the magnetic glueball masses ($m_{B}, m_{\phi}$) at $T=0$. A large reduction of the magnetic glueball masses, $m_{B}$ and $m_{\phi}$ have been observed, near the deconfinement temperature ($T_{d}$). The numerical estimates of vacuum expectation value of magnetic monopole condensate, the vector and scalar glueball masses at finite temperature for different coupling have been shown in table (\ref{Table:3}) respectively.
\begin{figure}
\centering
\includegraphics[width=.4\textwidth,origin=c,angle=360]{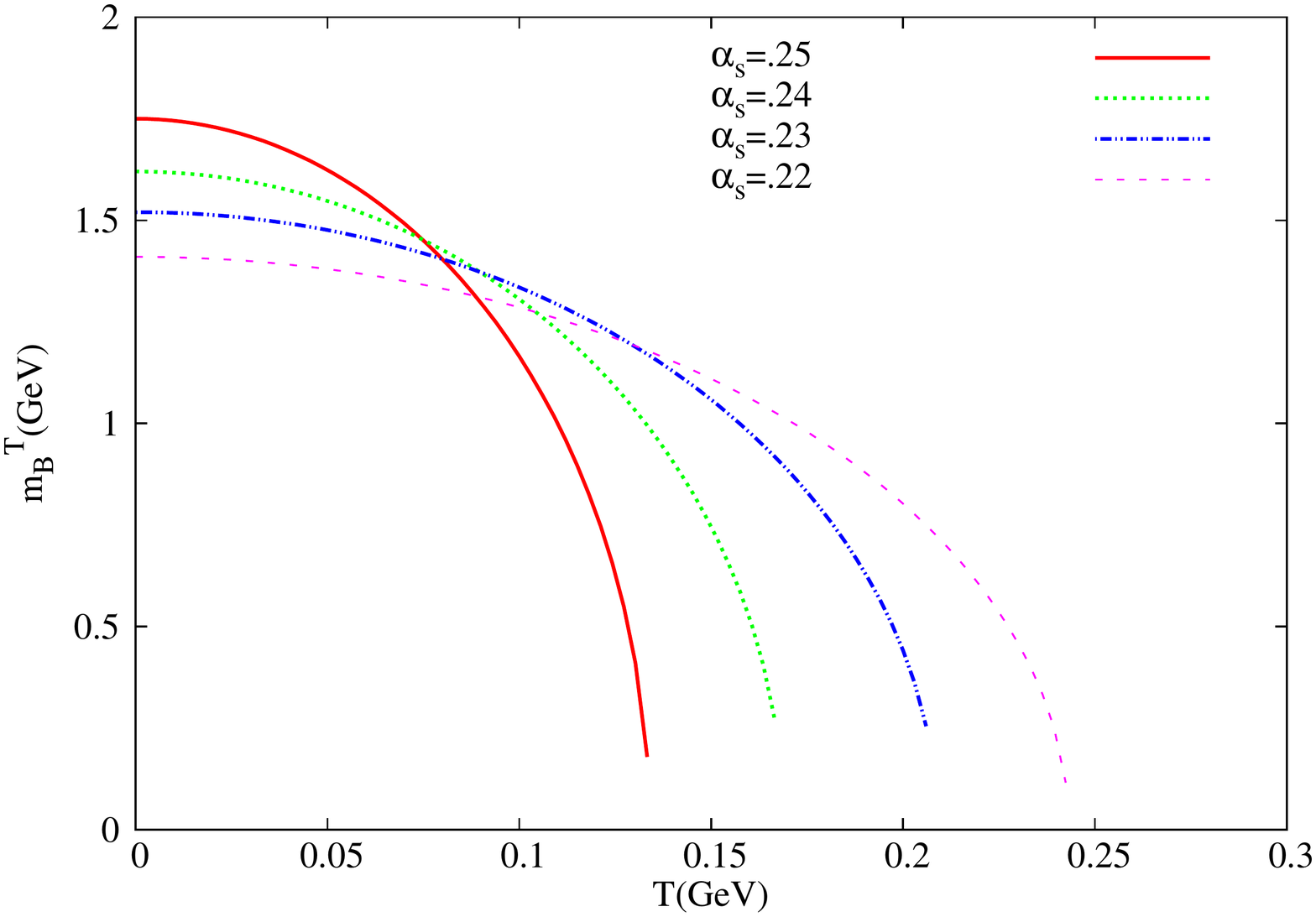}
\includegraphics[width=.4\textwidth,origin=c,angle=360]{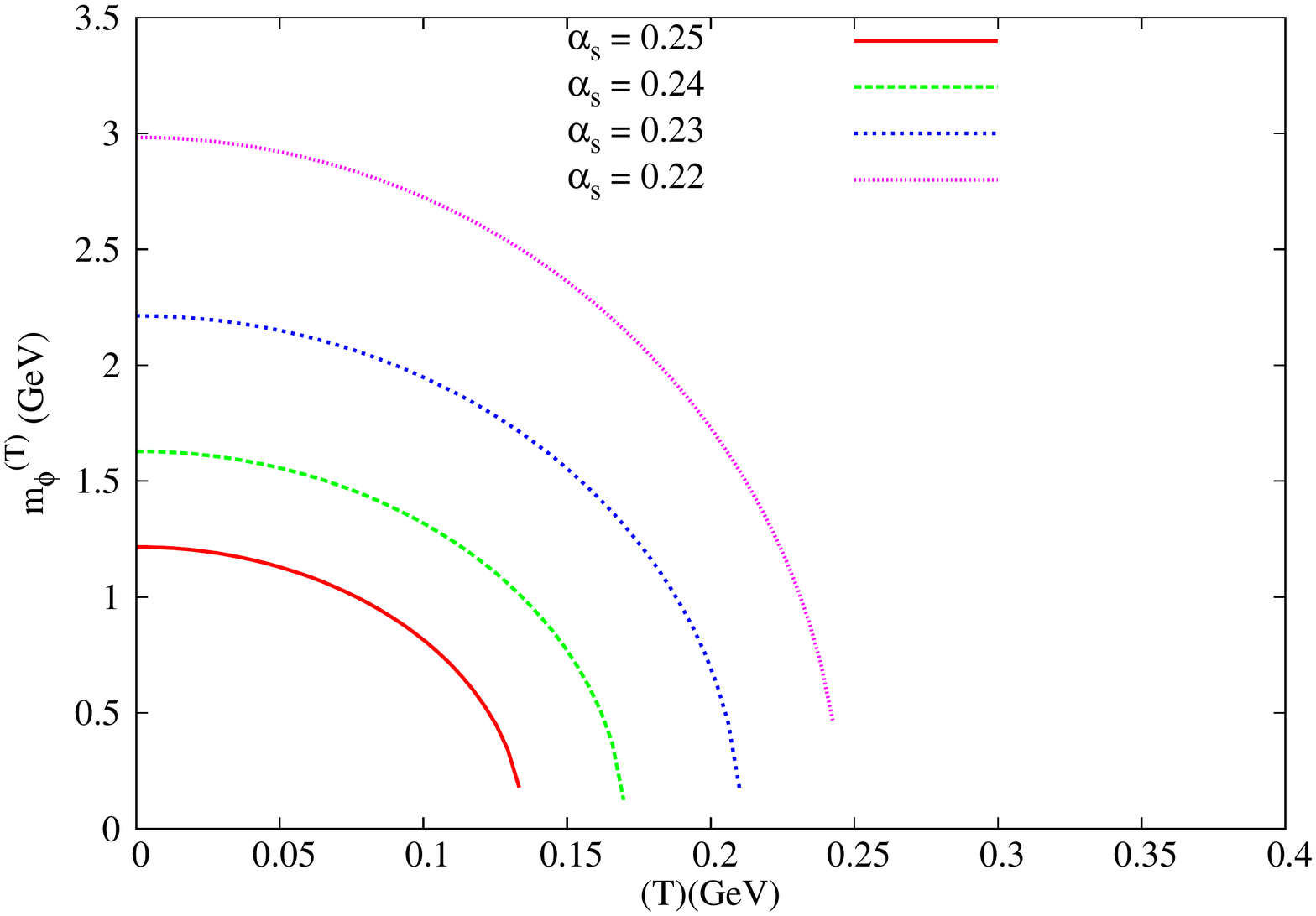}
\caption{(Color online.) Thermal response of vector and scalar glueball masses in $SU(3)$ dual QCD vacuum for $\alpha_s$ = $0.25$, $0.24$, $0.23$ and $0.22$ coupling respectively.}
 \label{Figure4}
\end{figure}
\begin{figure}
\centering
\includegraphics[width=.4\textwidth,origin=c,angle=360]{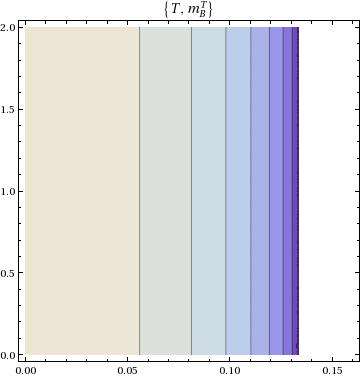}
\includegraphics[width=.4\textwidth,origin=c,angle=360]{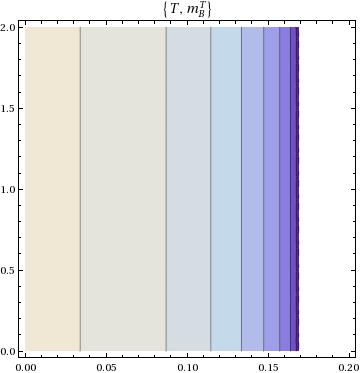}\\
\includegraphics[width=.4\textwidth,origin=c,angle=360]{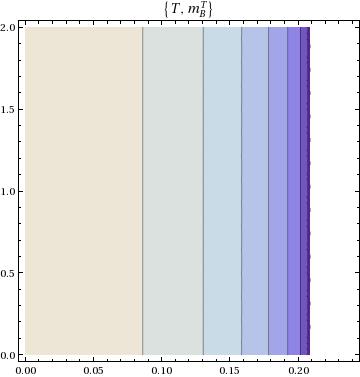}
\includegraphics[width=.4\textwidth,origin=c,angle=360]{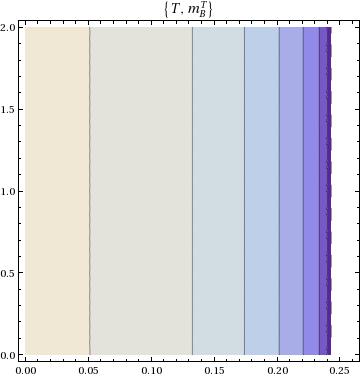}
\caption{(Color online.) Contour plot for the thermal response vector glueball mass with temperature in $SU(3)$ dual QCD vacuum for $\alpha_s$ = $0.25$, $0.24$, $0.23$ and $0.22$ coupling respectively. }
\label{Figure4}
\end{figure}
\begin{figure}
\centering
\includegraphics[width=.4\textwidth,origin=c,angle=360]{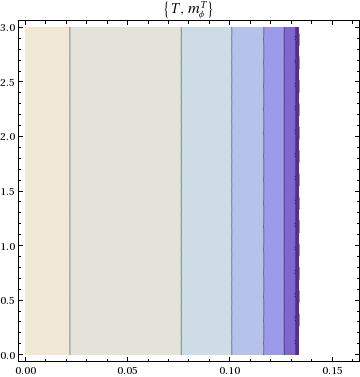}
\includegraphics[width=.4\textwidth,origin=c,angle=360]{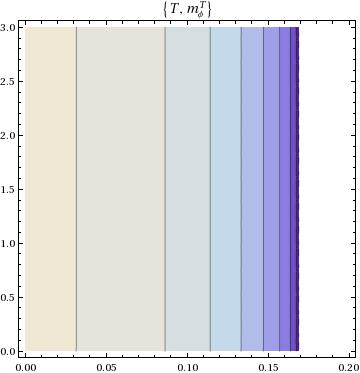}\\
\includegraphics[width=.4\textwidth,origin=c,angle=360]{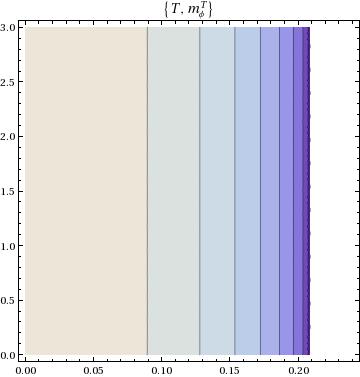}
\includegraphics[width=.4\textwidth,origin=c,angle=360]{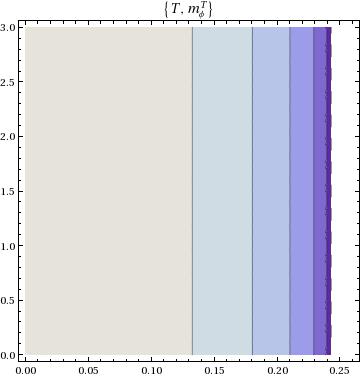}
\caption{(Color online.) Contour plot for the thermal response of the scalar glueball mass with temperature in $SU(3)$ dual QCD vacuum for $\alpha_s$ = $0.25$, $0.24$, $0.23$ and $0.22$ coupling respectively }
\label{Figure4}
\end{figure}
\begin{table}
\centering
\caption{The numerical estimate of the vacuum expectation value of the monopole field, vector glueball mass and scalar glueball mass at different values of coupling in $SU(3)$ dual QCD vacuum.}
\label{Table:3}
\begin{tabular*}{\columnwidth}{@{\extracolsep{\fill}}lllll@{}}
\hline
$\alpha_s$  & $<\phi>_{0(max)}^{(T)}(T=0)$ & $\bar{m}_{B(max)}^{(T)}(T=0)$ & $\bar{m}_{\phi(max)}^{(T)}(T=0)$ &$T_{d}$\\
&$(GeV)$&$(GeV)$&$(GeV)$&$(GeV)$\\
\hline
 0.22 & 0.176 &1.74 &1.21 & 0.135 \\
 \hline
 0.23 & 0.158 &1.63& 1.68& 0.170\\
 \hline
0.24 & 0.148 &1.53 & 2.16&0.210\\
\hline
0.25 & 0.134 &1.42 &2.89& 0.243\\ 
\hline
\end{tabular*}
\end{table}
\section{Summary and Conclusions} 
In the present paper a theoretical framework has been established to attain a low-energy effective theory of QCD approaching towards a first-principle derivation of confinement deconfinement phase transition at finite temperature. In fact, it has been manifested that an effective theory retrieved through this framework allows to hold both confinement and deconfinement transitions together on equal ground. The non-abelian dual superconductivity has been proposed in $SU(3)$ dual QCD formulation  for explaining the mechanism of color  confinement and gauge field decomposition to extract magnetic monopoles in a gauge invariant way. Based on the $SU(3)$ color gauge group, the color confinement has been investigated through the dual Meissner effect by specifying three main properties of the dual QCD vacuum. The first one allows the introduction of magnetic symmetry which best suited to describe the topological structure and thus monopole solutions of the underlying symmetry group. Consequently, the imposition of magnetic symmetry produces the magnetic condensation of dual QCD vacuum and guarantees the dual Meissner effect. Secondly, the dual Meissner effect confines the color electric flux carried by colored quark constituents producing string picture of hadrons. Thirdly, the physical spectrum of the colored quark constituent consists of the color-singlet combination owing to another residual symmetry called color reflection invariance. The existence of the scalar and vector magnetic glueball masses have been established at zero temperature which plays an important role in understanding confinement at zero temperature. Furthermore, we elucidate the mechanism of deconfinement transition phase transition in the pure gluon sector by describing the non-perturbative effective potential at various temperature in the imaginary-time formalism and occurrence of deconfinement temperature for phase transition. The disappearance of the dual Meissner effects and the breaking of the flux tube at high temperature have been demonstrated. The magnetic monopole condensate in the deconfinement 
phase transition has also been extracted and observed that the deconfinement phase transition is always associated with the disappearance of magnetic monopole condensate. This is the evidence that the deconfinement phase transition is caused by the disappearance of the non-Abelian dual superconductivity. The magnetic monopole condensate acts as an order parameter which signals the deconfinement phase transition, i.e.,  $\phi$ vanishes for high temperature $T>T_{d}$ signaling the deconfinement phase transition and is associated to the restoration of magnetic symmetry. The transition is discontinuous and the order is first order which is consistent with the Landau theory of phase transition and lattice gauge theory \cite{Hohenberg, Fukugita, Mizuki}. The deconfinement transition temperature has been extracted at $0.243 GeV$, $0.210 GeV$, $0.170 GeV$ and $0.135 GeV$ for the coupling $0.22$, $0.23$, $0.24$ and $0.25$ respectively in the $SU(3)$ dual QCD formulation. The temperature dependence of the magnetic glueball masses have been obtained from the minimization of the effective potential at finite temperature and Their disappearance enables us to easily understand the occurrence of the deconfinement phase transition at finite temperature. In such sense, the foundation of the transition allows an effective microscopic mechanism for deconfinement phase transition and as a consequence, the origin of the magnetic glueball masses at finite temperature has been more substantial and significant than that at zero temperature. 

\begin{acknowledgments}

Garima Punetha, is thankful to University Grant Commission (UGC), New Delhi, India, for the financial assistance under the UGC-RFSMS
research fellowship during the course of the study.

\end{acknowledgments}


\end{document}